\documentclass[aps,pra,groupedaddress,notitlepage]{revtex4-1}

\usepackage{amssymb}
\usepackage{amsmath}
\usepackage{amsfonts}
\usepackage{amsthm}
\usepackage{graphicx}
\usepackage{color}
\usepackage{bm}
\usepackage{cases}
\usepackage{url}
\usepackage{hyperref}

\begin{document}

\title{Structure functions and flatness of streamwise velocity in a turbulent channel flow}

\author{C. Granero-Belinch\'on$^{1,2}$} 
\affiliation{%
$^{1}$ Department of Mathematical and Electrical Engineering, IMT Atlantique, Lab-STICC, UMR CNRS 6285, 655 Av. du Technop\^ole, Plouzan\'e, 29280, Bretagne, France.}
\affiliation{$^{2}$ Odyssey, Inria/IMT Atlantique, 263 Av. G\'en\'eral Leclerc, Rennes, 35042, Bretagne, France.}
\email{Correspondence: carlos.granero-belinchon@imt-atlantique.fr}
\author{S. G. Roux$^{3}$}
\affiliation{$^{3}$ ENS de Lyon, CNRS, LPENSL, UMR5672, 69342, Lyon CEDEX 07, France}
\author{N. B. Garnier$^{4}$}
\affiliation{$^{4}$ CNRS, ENS de Lyon, LPENSL, UMR5672, 69342, Lyon CEDEX 07, France}

%


\begin{abstract}
In this article, we present a multiscale characterization of the streamwise velocity of a turbulent channel flow. We study the 2nd and 4th order structure functions and the flatness for scales ranging from the dissipative to the integral domains and for a wide range of distances to the walls spanning four distinct regions of the channel. We characterize the impact of the shear stress induced by the walls on these statistics. Far from the walls, in the {\em outer} layer, the impact of the boundaries on the flow is negligible and the flow statistics follow the Kolmogorov-Obukhov theory. In the {\em viscous}, {\em buffer} and {\em logarithmic} regions, the inertial domain can be split in two subdomains of scales with two different statistical behaviors. In the logarithmic region, the scaling of the structure functions agrees with the model of Davidson et al. 2006 but the scaling of the flatness seems to better correspond to the characterization of intermittency proposed by Kolmogorov and Obukhov in 1962. The structure functions and flatness of the streamwise velocity in the buffer and viscous regions are studied for the first time. We show the strong non-Gaussianity of the velocity flow at any scale in the viscous layer with strong intermittent events that may correspond to high shear-induced dissipation.
\end{abstract}

\maketitle 


\section{Turbulent flow in a channel}

Turbulence is a multiscale phenomenon in which the different scales interact non-linearly, producing a cascade of energy from large scales where energy is injected, to small scales where energy is dissipated~\cite{Richardson1921, Frisch1995}. For homogeneous and isotropic turbulence, this description results in the definition of three domains of scales~\cite{Richardson1921,Kolmogorov1991}. The integral domain where the energy is injected: the scales larger than the integral scale $L$. Then, the inertial domain where the energy cascades from large to smaller scales: scales smaller than $L$ and larger than the dissipative scale $\eta$. Finally, the dissipative domain where energy is dissipated: scales smaller than $\eta$. 

In wall-bounded flows, the walls produce a shear stress that decays with the distance to the walls and so introduce anisotropy and inhomogeneity in the flow. Eddy sizes are organized in function of the distance to the wall $\tilde{y}$ that appears as a new characteristic scale of the flow. The integral scale $L$, where energy is injected corresponds to the outer scale of the flow $\delta$, also called the boundary layer thickness or channel half-width. From classical theories of wall turbulence~\cite{Adrian2000,Marusic2019,Hwang2018}, we can consider four different regions of the flow in function of their distance to the wall~\cite{Jimenez2013}. First, the viscous and buffer regions that are the closest to the wall. Far from the walls, the outer layer where the small scales behave similarly to those of isotropic and homogeneous turbulence and the classical Kolmogorov theory of turbulence is recovered~\cite{Saddoughi1994}. While in the viscous and buffer regions, the shear dominates and the evolution of the turbulent scales is approximated as controlled by the mean flow~\cite{Jimenez2013a}, in the outer layer, non-linearity prevails. The transition between these two behaviors takes place in the logarithmic layer, where the smaller scales start to become independent of the shear. Despite their thinness, the viscous, buffer and logarithmic regions, that compose the inner layer, account for a major part of the total energy dissipation~\cite{Marusic2019} and so their study is of major interest. Moreover, in turbulent flows at low and intermediate Reynolds number, production of energy mainly happens in the inner layer, while at high Reynolds it occurs all along the boundary layer~\cite{Samie2018}. Close to the wall, different coherent structures cohabit: streamwise large scale motions or {\em superstructures} of the order of 10-25 $\delta$~\cite{Hutchins2007}, Hairpin vortices that born in the wall and cross the boundary layer~\cite{Theodorsen1952,Adrian2000}, vortex packets~\cite{Adrian2007} or long streaks along the streamwise direction in the viscous region~\cite{Cantwell1978} among others.

A classical statistical description of turbulence is based on structure functions of the velocity field $\vec{u}$~\cite{Kolmogorov1991} defined as:

\begin{equation}\label{eq:Sp}
S_p(l)= \left\langle |\delta_{l}\vec{u}(\vec{r}) |^{p} \right\rangle \,,
\end{equation} 

\noindent where $p$ indicates the order of the structure function, $\left\langle \, \right\rangle$ is the spatial average, $\vec{u}$ is the turbulent velocity field, $\vec{r}=(x,y,z)$ is the spatial position and $\delta_{l}\vec{u}(\vec{r})=\vec{u}(\vec{r}+l\vec{e}) - \vec{u}(\vec{r})$ is the velocity increment of length $l$ along the direction of the unitary vector $\vec{e}$. In particular, the second order structure function $S_2(l)$ is the variance of the increments and characterizes the distribution of energy across the scales of turbulence.

The irregular dissipation of energy in a turbulent flow is known as intermittency. It results in the probability density function of the turbulent velocity increments across scales to evolve from Gaussian at large scales to non-Gaussian with heavy tails at small scales. It also leads to the anomalous scaling of velocity increments~\cite{Frisch1995} that can be studied through the flatness: 
\begin{equation}
F(l) = \frac{S_4(l)}{3S_2(l)^2} \,. \label{eq:F}
\end{equation}

\noindent The flatness is the kurtosis of the increments; it characterizes the relative significance of extreme events as a function of the scale $l$ and so describes the non-Gaussianity of the velocity increments. A Gaussian distribution presents a kurtosis equal to $1$ and larger values of the kurtosis indicate larger tails of the distribution.

In this work, we focus on the study of the second and fourth order structure functions as well as of the flatness of the streamwise velocity of the flow, that we note $u_x$, in function of the distance to the wall $\tilde{y}$. Consequently, the spatial average is at fixed $\tilde{y}$ and we then write:

\begin{equation}\label{eq:Sp2}
S_{(p,\tilde{y})}(l)= \left\langle |u_x(x+l,y,z) - u_x(x,y,z)|^{p} \right\rangle_{y=\tilde{y}} \, ,
\end{equation} 

\noindent where $x$ is the streamwise direction, $y$ is the wall normal direction and $z$ is the spanwise direction. The streamwise velocity field can be written as $u_x(x,y,z)=U_x(y)+u_x'(x,y,z)$ where $U_x(y)=\left\langle u_x(x,y,z) \right\rangle_{y=\tilde{y}}$ is the mean velocity field that depends on the distance to the wall and $u_x'(x,y,z)$ represents the velocity fluctuations.

\subsection{The outer layer: fully developed turbulence far frow the walls}

Kolmogorov 1941 description of turbulence states that at very high Reynolds number (fully developed turbulence) and far away from boundaries, the turbulent flow becomes statistically homogeneous and isotropic and is scale-invariant in the inertial domain~\cite{Kolmogorov1991, Frisch1995}. This implies the following scaling for the structure functions $S_p(l)$:

\begin{equation}\label{eq:Spout}
S_p(l)= \left\langle |\delta_{l}u_x|^{p} \right\rangle \sim l^{\zeta_p} 
\end{equation} 

\noindent for scales $\eta<l<\delta$ and where $\zeta_p$ is the scaling exponent function that varies among the different models of turbulence~\cite{Dubrulle2019,Paladin1987}. A Taylor expansion of the scaling exponent function allows us to define the log-cumulants~\cite{Delour2001}:

\begin{equation}
\zeta_p = c_1 p - c_2 \frac{p^2}{2!} + c_3 \frac{p^3}{3!} ...
\end{equation}

\noindent While $c_1$ characterizes the global roughness of the velocity, $c_2$ is the intermittency parameter and models the local heterogeneity of this roughness which is linked to the existence of extreme events in the velocity increments.

The Kolmogorov 1941 model considers a linear behavior of the scaling exponent $\zeta_p=p/3$, \textit{i.e.} the turbulent velocity field is self-similar with $c_1=1/3$. On the other hand, the log-normal model~\cite{Kolmogorov1962,Obukhov1962} considers a quadratic form of the scaling exponent $\zeta_p= c_1 p - c_2\frac{p^2}{2}$. From the $4/5$-law of Kolmogorov $c_1=\frac{1}{3}+\frac{3}{2}c_2$ and $c_2=0.025$ from fit of experimental data~\cite{Chevillard2012,Paladin1987,Dubrulle2019,GraneroBelinchon2018}. More complex models exist that consider log-cumulants of order higher than two~\cite{She1994}. 

In the outer layer, the structure functions present three different behaviors depending on the scale of analysis, that correspond respectively to the dissipative, inertial and integral domain of scales. Considering the log-normal model for the scales in the inertial domain, the structure function of order $p$ is:

\begin{numcases}{S_{p}(l) \sim }
    l^{p} &for  $l<\eta$ \label{eq:Spdis}\\
    l^{c_1p-c2\frac{p^2}{2}} &for $\eta<l<\delta$ \label{eq:Splog}\\
    \left(2 \left\langle u_x'^{\, 2} \right\rangle \right)^{p/2} &for $l > \delta$
\end{numcases}

\noindent with $\delta$ the channel half width that corresponds to the integral scale $L$. In the dissipative domain, the structure functions behave as $l^p$ corresponding to a smooth velocity field~\cite{Bachelor1951} and in the integral domain they reach a plateau indicating decorrelation of the velocity field.

The corresponding flatness is:

\begin{numcases}{\log(F(l)) \sim }
    g(l) &for  $l<\eta$ \\
    l^{-4c2} &for $\eta<l<\delta$  \label{eq:Fouter}\\
    0 &for $l > \delta$
\end{numcases}

\noindent where $g$ is a generic function with a plateau for scales $l\ll\eta$ and a steep decrease when $l$ approach $\eta$~\cite{Chevillard2005}, and $\log(F(l))=0$ in the integral domain implies Gaussian distribution at these scales. Throughout this article $\log$ denotes natural logarithm. 

\subsection{The logarithmic layer}

The logarithmic layer is the zone between the shear-dominated region and the outer region where shear is negligible. It is defined by the linear dependence of the characteristic length scale of coherent structures with the distance to the wall~\cite{Millikan1938,Tennekes1972}. Experimental and numerical observations agree with the mean and variance of the streamwise velocity of the flow depending logarithmically on the distance from the wall~\cite{Townsend1961,Kunkel2006,Hultmark2012,Marusic2013,Silva2015,Dubrulle2024}:

\begin{eqnarray}
\left\langle u_x \right\rangle_{\tilde{y}} &=& \frac{1}{\kappa} \log(\tilde{y}) + A \label{eq:logMP}\\
\left\langle u_x'^{\,2} \right\rangle_{\tilde{y}} &\sim& \log(\delta/\tilde{y})
\end{eqnarray}

\noindent where $A$ is a constant that depends on the flow and the Karman constant $\kappa \approx 0.4$ is considered universal~\cite{Jimenez2013}.

In the logarithmic layer, the shear is considered to be dominant at scales larger than a given transition length, while it is considered to be negligible at smaller scales~\cite{Toschi1999}. Recent works showed, for the streamwise velocity in the logarithmic region, the behavior of the even-order structure functions $S_p(l)$ in function of the distance to the walls $\tilde{y}$, pointing out the existence of this transition length that is of the order of $\kappa\tilde{y}$. Thus, four domain of scales with four different behaviors are observed in the structure functions~\cite{Silva2015}:

\begin{numcases}{S_{(p,\tilde{y})}(l) \sim }
    l^{p} &for  $l<\eta$ \label{eq:Sdis}\\
    C_pl^{p/3} &for $\eta<l<\tilde{y}$ \label{eq:SK41}\\
    \left( E_p+D_p\log(l/\tilde{y})\right)^{p/2} &for $\tilde{y}<l \ll \delta$ \label{eq:Slog}\\
    G_p \left(2 \left\langle u_x'^{\, 2} \right\rangle_{\tilde{y}} \right)^{p/2} &for $l > \delta$
\end{numcases}

\noindent where $C_p$, $E_p$, $D_p$ and $G_p$ are constants. As in the outer layer, the behavior in $l^p$ for smaller scales below the Kolmogorov dissipative scale indicates that the velocity is perfectly smooth at these scales, and the plateau in the integral domain implies decorrelation of the velocity field. The inertial domain is divided in two subdomains. The scales between $\eta$ and $\tilde{y}$ present a scaling in $l^{p/3}$ in agreement with K41 theory, \textit{i.e.} non-linearity governs, while for scales $l$ larger than the distance to the wall $\tilde{y}$ and smaller than $\delta$, the shear dominates and the energy density is hypothesized to scale with $1/l$~\cite{Davidson2006a}. The second order structure function $S_2(l)$ characterizes the cumulative energy of scales $l$ and smaller. Then, the logarithmic behavior at scales $\tilde{y}<l \ll \delta$ described in (\ref{eq:Slog}) appears from the integration of the energy density~\cite{Davidson2006a}.

These results have been illustrated on velocity measurements at different distances $\tilde{y}$ from different experimental set-ups at different Reynolds numbers~\cite{Davidson2014, Silva2015}. Equation (\ref{eq:SK41}) can be easily generalized to take into account intermittency by considering the log-normal model instead of K41 just like in equation (\ref{eq:Splog}). Finally, intermittency can be studied finely with flatness that also presents four different behaviors depending on the scales:

\begin{numcases}{\log(F_{\tilde{y}}(l)) \sim }
    g(l) &for  $l<\eta$  \\
    l^{-4c2} &for $\eta<l<\tilde{y}$ \label{eq:FK41}\\
    \left( \frac{E_4+D_4\log(l/\tilde{y})}{E_2+D_2\log(l/\tilde{y})}\right)^{2} &for $\tilde{y}<l \ll \delta$ \label{eq:Flog}\\
    0 &for $l > \delta$
\end{numcases}

The flatness in the dissipative and integral domains behaves as in the outer layer, but in the logarithmic region the inertial domain is now divided into two subdomains of scales with different behaviors of the flatness.

\subsection{The viscous and buffer regions}

In the {\em viscous} layer, the closest to the wall, the mean velocity of the flow behaves as:

\begin{equation} \label{eq:viscMP}
\left\langle u_x \right\rangle_{\tilde{y}} \sim \tilde{y}
\end{equation}

The {\em buffer} layer is the region of transition from $\left\langle u_x \right\rangle_{\tilde{y}} \sim \tilde{y}$ in the viscous region to $\left\langle u_x \right\rangle_{\tilde{y}} \sim \log(\tilde{y})$ in the logarithmic region. While in the logarithmic and outer layers structures of multiple scales coexist, there should be only single-length structures close to the wall, because the energy injection and dissipation occur at similar scales~\cite{Jimenez2013,Jimenez2013a,Hwang2018}. From the best of our knowledge, a detailed study of the structure functions and flatness of the streamwise velocity has not been performed for $\tilde{y}$ in these shear-dominated regions. 

\subsection{Contributions}

We characterize the behavior of the second and fourth order structure functions as well as the flatness of the streamwise velocity in a turbulent channel flow in function of the distance to the walls. We study an ensemble of distances belonging to four different regions: viscous, buffer, logarithmic and outer. We do so on Direct Numerical Simulation (DNS) of a turbulent channel flow from the open access turbulence database of Johns Hopkins University~\cite{Lee2015}. In the outer layer, we illustrate the perfect agreement of $S_2(l)$, $S_4(l)$ and $F(l)$ with the log-normal model of turbulence. In the logarithmic layer, we recover the results from~\cite{Davidson2014, Silva2015} for $S_{(2,\tilde{y})}(l)$ and $S_{(4,\tilde{y})}(l)$ and we illustrate the intermittent nature of the velocity flow at all scales $l<\delta$ with the flatness. In the viscous and buffer regions, we still find a domain of scales $l<\eta$ where dissipation dominates, but, even at small scales, we don't observe Kolmogorov scaling. Instead, we find a transition scale $d$ defined as three times the thickness of the viscous and buffer region. For scales $\eta<l<d$ the structure functions scale as $S_p(l)=\left(E_p+D_p\log(l/d)\right)^{p/2}$ while for scales $d<l<\delta$ the structure functions behave as $S_p(l)=\left(K_p+L_p\log(l/d)\right)^{p/4}$. In the buffer region the flatness tends to the Gaussian value at large scale and in the viscous region it tends to a non-Gaussian value of $F(l)=0.41$. 

This article is organized as follows. In section~\ref{sec:data}, we present the data from DNS used in this study. In section~\ref{sec:results}, we illustrate the behavior of the second and fourth order structure functions and flatness of the streamwise velocity for the four different regions of the channel flow. In section~\ref{sec:conclusion}, we present the main conclusions and some perspectives. 

\section{Johns Hopkins University Channel5200 turbulent velocity dataset}\label{sec:data}

\begin{figure}[ht]
 \includegraphics[width=0.8\linewidth]{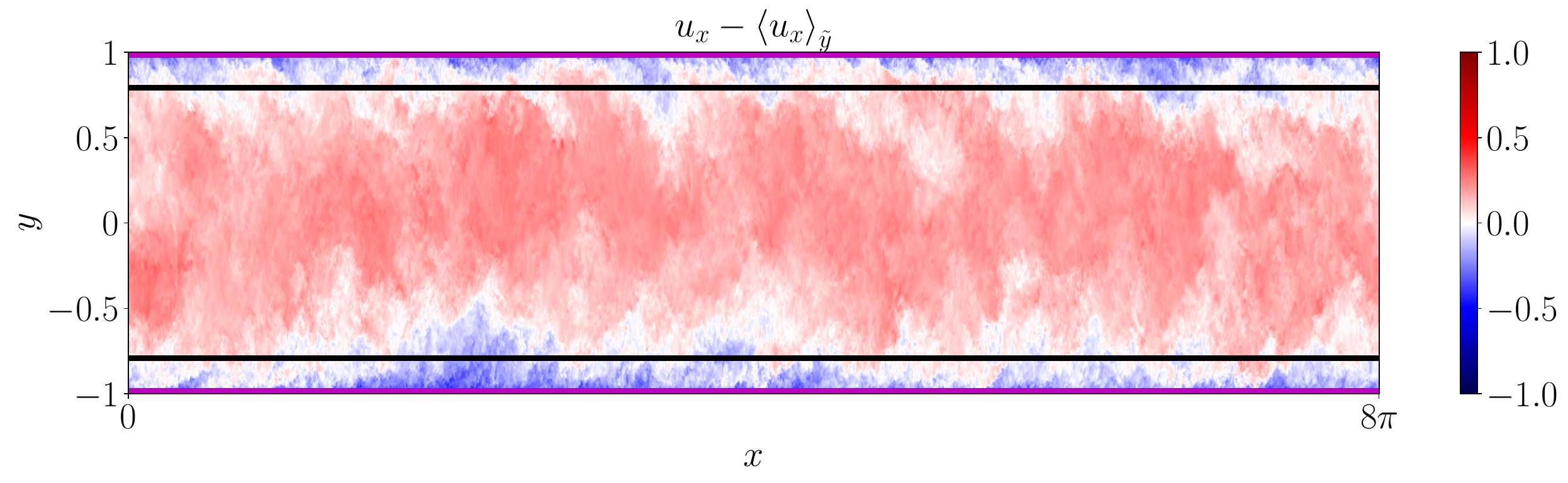}
 \centering
 \caption{Two-dimensional transect of streamwise $u_x$ from the 3-d velocity field from JHU Channel turbulent flow Direct Numerical Simulation~\cite{Lee2015}. The $(x,y)$ coordinates span the full spatial region $([0, 8\pi], [-1, 1])$ and $z=\frac{3\pi}{2}$ is fixed. The horizontal colored lines indicate the borders between: in magenta the buffer and logarithmic regions ($y\in\{-0.981,0.981\}$), in black the logarithmic and the outer layers ($y\in\{-0.792,0.792\}$).}\label{fig:channel5200}
\end{figure}

\begin{figure}[ht]
 \includegraphics[width=0.6\linewidth]{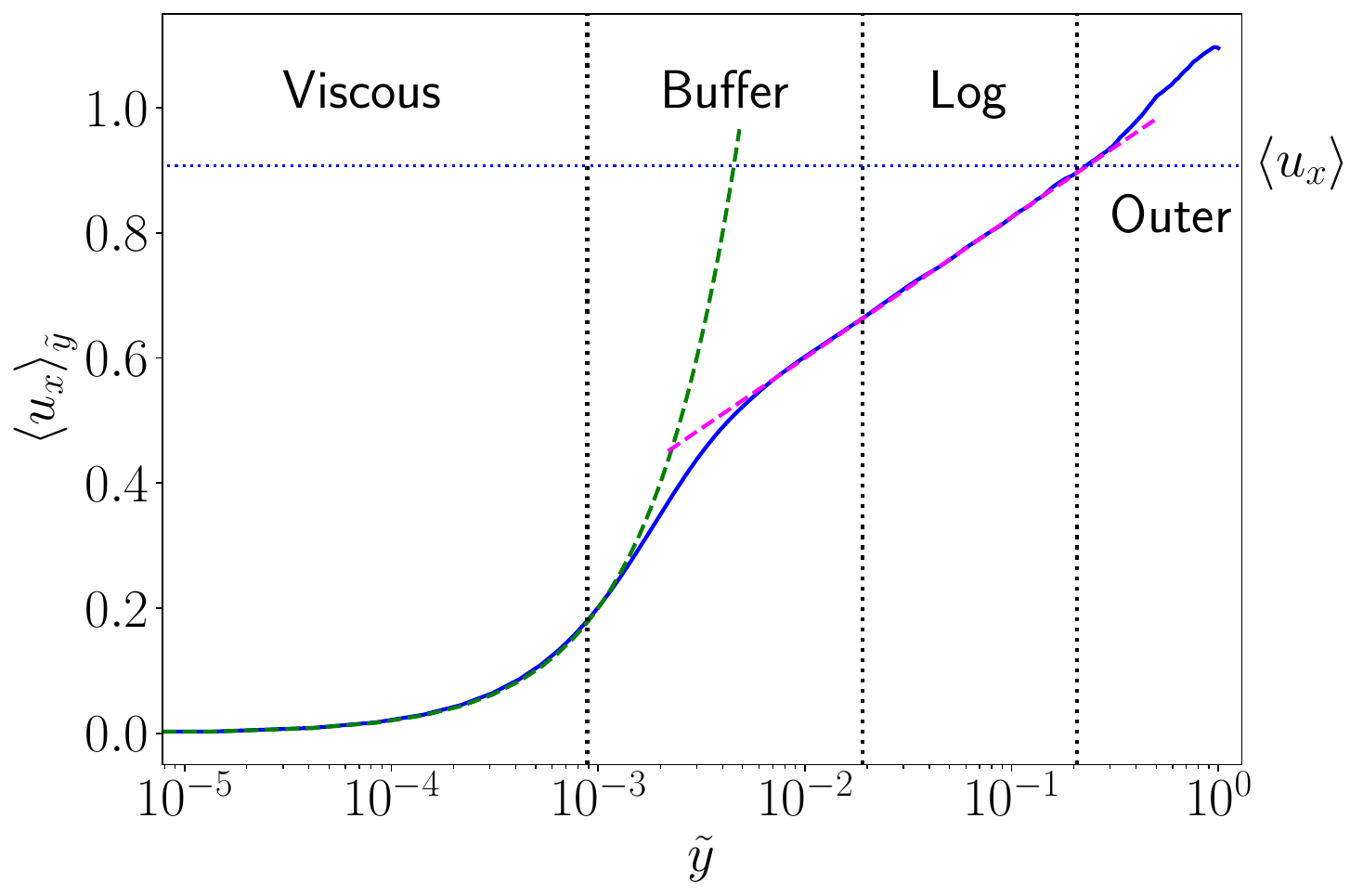}
 \centering
 \caption{Mean streamwise velocity profile in function of the logarithm $\log_{10}$ of the distance to the wall. The green dashed line corresponds to the expected behavior in the viscous region, eq.(\ref{eq:viscMP}), and the magenta dashed line corresponds to the expected behavior in the logarithmic region, eq.(\ref{eq:logMP}). The three vertical dotted lines indicate the demarcations between regions: viscous and buffer, buffer and logarithmic and logarithmic and outer respectively. The horizontal dotted blue line corresponds to the mean velocity of the streamwise velocity of the flow $\left\langle u_x \right\rangle$.}\label{fig:channel5200MP}
\end{figure}

We study here a 2D transect of the longitudinal component $u_x$ of a three dimensional turbulent velocity field obtained from a DNS of a turbulent channel flow with periodic boundary conditions in the $x$ and $z$ directions, and with no-slip boundary conditions for $y=\pm1$~\cite{Lee2015}. The domain of the simulation is $(x,y,z) \in [0, 8\pi[ \times [-1, +1] \times [0, +3\pi[$. The 2D transect is at $z=\frac{3\pi}{2}$; it is an image of size $10240 \times 1536$ pixels in the $x$ and $y$ directions. A pixel corresponds to $\pi/1280$ along $x$ and non-regular sampling along $y$, see figure~\ref{fig:channel5200}. The friction-velocity Reynolds number of the flow is $R_{\tau}=5200$, and the Reynolds number in the bulk is $R_b=1.25 \times 10^{5}$. Consequently, we consider that the flow exhibits fully developed turbulence regime. The forcing is stationary and acts only at large scale, so the inhomogeneity and anisotropy of the flow are only induced by the boundaries. The anisotropy of the velocity field is two-fold. First, there is no mean flow in the $y$ and $z$ directions while there is a mean velocity along the $x$ direction (streamwise direction). Second, the mean velocity profile $\left\langle u_x \right\rangle_{\tilde{y}}$ is inhomogenous due to the existence of boundary layers around $y=\pm 1$, see figure~\ref{fig:channel5200MP}. These boundaries induce the existence of four different regions. The viscous region exists for $y \in [-1,-0.999] \cup [0.999,1]$, the buffer region for $y \in [-0.999,-0.981] \cup [0.981,0.999]$, the logarithmic region for $y \in [-0.981,-0.792] \cup [0.792,0.981]$ and the outer layer for $y \in [-0.792,0.792]$. This corresponds to a logarithmic region beginning at $y_+=98$ and finishing at $y_+=1075$~\cite{Lee2015}, where $y_+=\tilde{y}/\delta_{\nu}$ is a normalized distance to the wall and $\delta_{\nu}=1.9283 \times10^{-4}$ is the viscous length scale of the flow. The borders between the different layers are illustrated by continuous horizontal lines in figure~\ref{fig:channel5200}. The mean velocity profile $\left\langle u_x \right\rangle_{\tilde{y}}$ is represented in function to the distance to the wall $\log_{10}(\tilde{y})$ in figure~\ref{fig:channel5200MP}, illustrating the linear behavior (\ref{eq:viscMP}) in the viscous region (green dashes), the logarithmic behavior (\ref{eq:logMP}) in the logarithmic region (magenta dashes), and the transition in the buffer layer. 


To study the structure functions of the streamwise velocity for $\tilde{y}$ distances from the viscous region to the outer one, we normalized the streamwise velocity individually per $\tilde{y}$ position. In fact, for each $\tilde{y}$, the streamwise velocity is standardized using its mean and standard deviation both estimated at the corresponding $\tilde{y}$. This allows to avoid the strong evolution of the mean and the variance profiles of the streamwise velocity with the distance to the wall.


\section{Results}
\label{sec:results}

\begin{figure}[ht!]
\includegraphics[width=\linewidth]{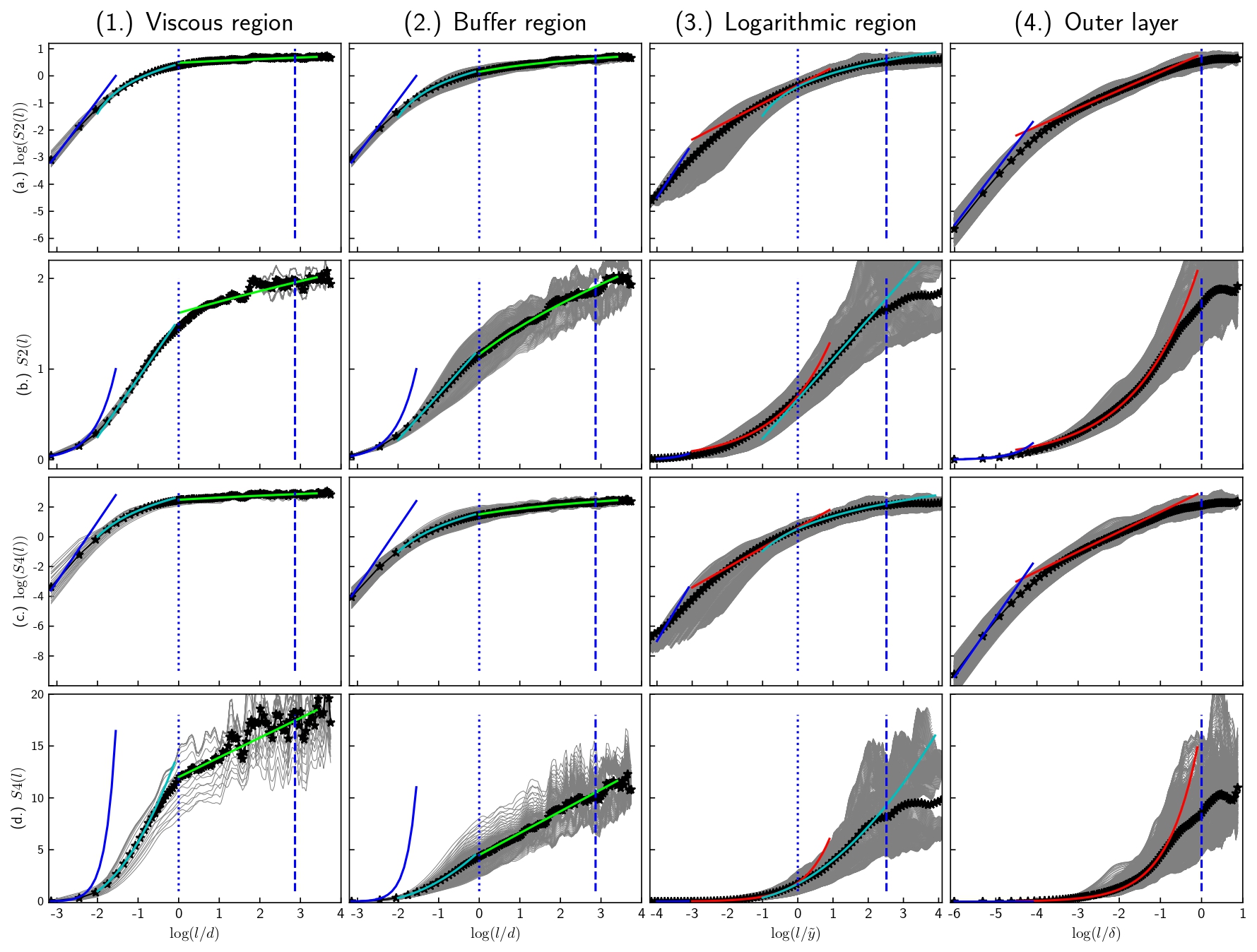}
 \centering
\caption{\textbf{Second and fourth order structure functions.} a. $\log(S_2(l))$, b. $S_2(l)$, c. $\log(S_4(l))$ and d. $S_4(l)$ for the viscous region (1.), buffer region (2.), the logarithmic region (3.) and the outer layer (4.) in function of $\log(l/d)$, $\log(l/\tilde{y})$ and $\log(l/\delta)$ respectively. Gray lines represent the structure function for a given $y$ position within each specific region. The black lines are the averages over $y$. In the outer region, the blue and red curves correspond respectively to equations (\ref{eq:Spdis}) and (\ref{eq:Splog}). In the logarithmic region, the blue, red and cyan curves are the theoretical behavior from equations (\ref{eq:Sdis}), (\ref{eq:SK41}) and (\ref{eq:Slog}). In the buffer and viscous regions, the blue, cyan and green lines indicate the behavior from equations (\ref{eq:Sdisvisbuf}), (\ref{eq:Slog1visbuf}) and (\ref{eq:Slog2visbuf}). The vertical blue dotted line in the viscous and buffer regions represent $d$ and in the logarithmic region $\tilde{y}$. The vertical blue dashed lines in (1.), (2.), (3.) and (4.) represent $\delta$.}\label{fig:S2S4}
\end{figure}

Figure~\ref{fig:S2S4} illustrates the behavior of $\log(S_{(2,\tilde{y})}(l))$ (a), $S_{(2,\tilde{y})}(l)$ (b) $\log(S_{(4,\tilde{y})}(l))$ (c) and $S_{(4,\tilde{y})}(l)$ (d) as the scale $l$ is varied, for $\tilde{y}$ positions in the viscous region (1.), in the buffer region (2.), in the logarithmic region (3.) and in the outer layer (4.). The scale of analysis is normalized differently depending on the studied region. First, in the viscous and buffer regions we use $l/d$ with $d$ being three times the thickness of the viscous and buffer regions. Indeed, following~\cite{Jimenez2013} the near-wall region should have a single characteristic scale and so considering a normalization by a single scale $d$ seems appropriate. Second, in the logarithmic region we follow the approach of~\cite{Davidson2006,Davidson2006a,Davidson2009,Davidson2014,Silva2015} and use $l/\tilde{y}$. Finally, in the outer layer we use $l/\delta$, \textit{i.e.} the half-width of the channel $\delta$, that corresponds to the integral scale. 

In the outer layer, figure~\ref{fig:S2S4} (4.), the behaviors of the second and fourth order structure functions correspond perfectly to the ones of fully developed homogeneous and isotropic turbulence: $S_p(l)$ presents a plateau at scales $l\geq \delta$, a Kolmogorov power law $l^{p/3}$ at scales $\eta<l<\delta$ and a exponential decrease $l^p$ in the dissipative domain $l<\eta$. The blue and red curves correspond respectively to the known behaviors for the dissipative and inertial domains of scales from~\cite{Bachelor1951}. 

In the logarithmic region, figure~\ref{fig:S2S4} (3.), the second and fourth order structure functions behave accordingly to the model from~\cite{Davidson2014,Silva2015}. At scales $l<\eta$ dissipation governs and the structure functions tend to $l^p$ scaling (blue line). The inertial domain is now divided in two subdomains: first, $\eta<l<\tilde{y}$ on which we recover the K41 scaling $l^{p/3}$ (red line), and second $\tilde{y}<l<\delta$ on which we recover $\left(E_p+D_p\log(l/\tilde{y})\right)^{p/2}$ (cyan line). The constants $E_p$ and $D_p$ measured in this region are shown in table~\ref{tab:1}. We recall that the used normalization of the velocity field depends on $\tilde{y}$ and so leads to $D_p$ and $E_p$ constants that do not correspond to the values of~\cite{Davidson2009,Davidson2014,Silva2015} supposed universal when $u_+=u_x/u_{\tau}$ is studied with $u_{\tau}=4.14872 \times 10^{-2}$ the mean friction velocity of the flow. 

In the buffer and viscous regions, figures~\ref{fig:S2S4} (2.) and~\ref{fig:S2S4} (1.) respectively, four different domains of scales appear: a dissipative domain where the structure functions tend to $l^p$ (blue line), a first inertial domain with $\left(E_p+D_p\log(l/d)\right)^{p/2}$ (cyan line), a second inertial domain with $\left(K_p+L_p\log(l/d)\right)^{p/4}$ and a plateau for scales $l>\delta$. Strong differences of $\left\langle u_x'^{\, 2} \right\rangle_{\tilde{y}}$ with $\tilde{y}$, prevent from the normalization of the analysis scale used in the logarithmic region, \textit{i.e.} $\log(l/\tilde{y})$, that does not allow to superpose the observations for different $\tilde{y}$.

From figures~\ref{fig:S2S4} (2.) and~\ref{fig:S2S4} (1.), we propose the following model for the structure functions in the viscous and buffer regions:

\begin{numcases}{S_{p}(l) \sim }
    l^{p} &for  $l<\eta$ \label{eq:Sdisvisbuf}\\
    \left( E_p+D_p\log(l/d)\right)^{p/2} &for $\eta<l<d$ \label{eq:Slog1visbuf}\\
    \left( K_p+L_p\log(l/d)\right)^{p/4} &for $d < l < \delta$ \label{eq:Slog2visbuf}\\
    G_p \left(2 \left\langle u_x'^{\, 2} \right\rangle_{\tilde{y}} \right)^{p/2} &for $l > \delta$
\end{numcases}

\noindent which leads to the following model for the flatness:

\begin{numcases}{\log(F(l)) \sim }
    g(l) &for  $l<\eta$ \label{eq:Fdis} \\
    \left( \frac{E_4+D_4\log(l/d)}{E_2+D_2\log(l/d)}\right)^{2} &for $\eta<l<d$ \label{eq:F1visbuf}\\
     \frac{K_4+L_4\log(l/d)}{K_2+L_2\log(l/d)} &for $d<l \ll \delta$ \label{eq:F2visbuf}\\
    0 &for $l > \delta$
\end{numcases}

\noindent where the obtained constants are indicated in table~\ref{tab:1}.

\begin{table*}
 \caption{Constants of the analytical expressions of the second and fourth order structure functions in the logarithmic, buffer and viscous regions. $E_p$ and $D_p$ are taken from (\ref{eq:Slog}) for the logarithmic region and from (\ref{eq:Slog1visbuf}) for the buffer and viscous regions; $K_p$ and $L_p$ are taken from (\ref{eq:Slog2visbuf}).}\label{tab:1}
 \begin{center}
\begin{tabular}{ |p{2cm}|p{3cm}|p{3cm}|p{3cm}|}
 \hline
 Constants & Logarithmic Region & Buffer Region  & Viscous Region  \\ \hline
 $E_2$     & 0.67         & 1.22    & 1.55 \\ \hline
 $D_2$     & 0.44         & 0.50    & 0.65 \\ \hline
 $E_4$     & 1.31         & 2.20    & 3.80 \\ \hline
 $D_4$     & 0.69         & 0.80    & 1.40 \\ \hline
 $K_2$     & --           & 1.35    & 2.62 \\ \hline 
 $L_2$     & --           & 0.80    & 0.42 \\ \hline
 $K_4$     & --           & 4.50    & 12.00 \\ \hline
 $L_2$     & --           & 2.10    & 1.90 \\ \hline
\end{tabular} 
\end{center}
\end{table*}

In these regions, all the scales $l$ larger than the dissipative scale $\eta$ and smaller than the integral scale $\delta$ are governed by the shear. However, for scales smaller than $d$, we find the same behavior as in the shear-dominated scales of the logarithmic region, while for scales larger than $d$ the fourth order structure function scales linearly instead of the second order one. This corresponds to the energy density scaling as $l^{-1}$ for scales $\eta<l<d$ and $l^{-1}/\sqrt{\log(l)}$ for scales $d<l<\delta$.

\begin{figure}[ht!]
\includegraphics[width=\linewidth]{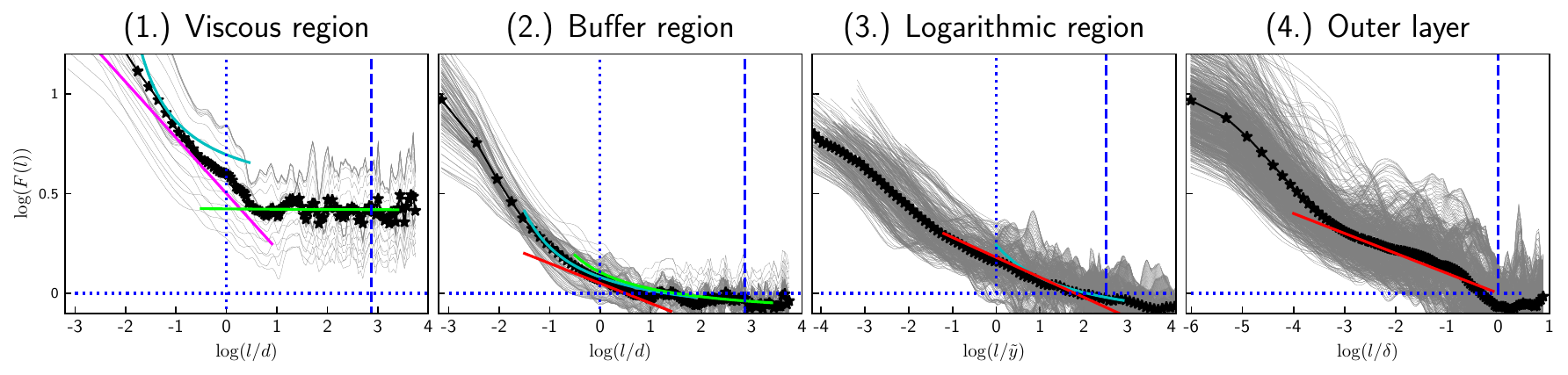}
 \centering
\caption{\textbf{Flatness.} $\log(F(l))$ for the viscous (1.), the buffer (2.), the logarithmic (3.) and the outer (4.) regions in function of $\log(l/d)$, $\log(l/\tilde{y})$ and $\log(l/\delta)$ respectively. Gray lines represent the flatness for a given $y$ position within each specific region. The black lines are the averages over $y$. In the outer region, the red line corresponds to equation (\ref{eq:Fouter}). In the logarithmic region, the red and cyan curves are the theoretical behavior from equations (\ref{eq:FK41}) and (\ref{eq:Flog}). For the buffer region the cyan and green lines indicate the behavior from (\ref{eq:F1visbuf}) and (\ref{eq:F2visbuf}) respectively and the red line correspond to (\ref{eq:Fouter}). In the viscous region, the green line is a plateau at $\log(F(l))=0.41$ corresponding to eq.(\ref{eq:F2visbuf}), the cyan line comes from (\ref{eq:F1visbuf}) and the magenta line has a slope of $-0.19$. The vertical blue dotted lines in the viscous and buffer regions represent $d$ while in the logarithmic region represents $\tilde{y}$. The vertical blue dashed lines in (1.), (2.), (3.) and (4.) represent $\delta$.}\label{fig:flatness}
\end{figure}

Figure~\ref{fig:flatness} shows the evolution of the flatness in the viscous (1.), buffer (2.), logarithmic (3.) and outer region (4.) in function of $\log(l/d)$, $\log(l/\tilde{y})$ and $\log(l/\delta)$ respectively. In all four cases, the flatness presents a plateau for scales $l>\delta$ and a very steep increase when the scale decreases in the dissipative domain $l<\eta$~\cite{Chevillard2012}. 

In the outer layer, the flatness in the inertial domain decreases linearly when the scale increases, with a slope of $-0.1=-4c_2$ (red line) in agreement with the log-normal model~\cite{GraneroBelinchon2018, Chevillard2012}. In the logarithmic region, the same linear behavior correctly matches the flatness in the inertial domain $\eta<l<\delta$. Although eq.(\ref{eq:Flog}) also seems to correctly model the flatness for scales $\tilde{y}<l<\delta$, the linear behavior with slope $-0.1$ better follows the observations. 

In the buffer region, the flatness obtained with equations (\ref{eq:F1visbuf}) and (\ref{eq:F2visbuf}) are plotted respectively in cyan and green together with the linear behavior from (\ref{eq:FK41}) in red. Gaussianity is reached at scales smaller than $\delta$, the linear behavior with slope $-0.1$ seems to correctly model some scales of the inertial domain ($-0.5<\log(l/d)<1$), and both cyan and green curves are close to the observed flatness. It is difficult to conclude which model better fits the observations. In the viscous region, the flatness does not reach Gaussian values at large scale, instead a plateau at $\log(F(l))=0.41$ appears illustrating the non-Gaussianity with heavy tails of the distribution of large-scale increments in this region. Furthermore, a steep linear decrease with slope $-0.19$ of the flatness is observed when the scale increase for scales $\log(l/d)<0.5$. The flatness from eq.(\ref{eq:F1visbuf}) is plotted in cyan for this region but it does not correctly model the observed flatness.

\section{Discussion and Conclusion} \label{sec:conclusion}

Figures~\ref{fig:S2S4} and~\ref{fig:flatness} illustrate four different behaviors for the structure functions of the streamwise velocity in a turbulent channel flow depending on the distance to the wall. Four different regions are observed that correspond to the: viscous, buffer, logarithmic and outer layers as defined by classical studies of turbulent channel flows~\cite{Jimenez2013,Davidson2009}. In the outer layer, we recover the behavior described by the Kolmogorov-Obukhov theory for both the structure functions and the flatness~\cite{Kolmogorov1962, Obukhov1962, GraneroBelinchon2018, Paladin1987, Chevillard2012}, indicating that at distance $y_+>1075$ the impact of the walls on the flow is negligible. In the logarithmic region $98<y_+<1075$, the results on the structure functions are in agreement with recent observations~\cite{Davidson2014,Silva2015}: the inertial domain of scales may be decomposed into scales $l<\tilde{y}$ where Kolmogorov-Obukhov theory prevails, \textit{i.e,} the non-linear advection governs the dynamics, and into scales $l>\tilde{y}$ where the shear stress dominates~\cite{Jimenez2013}. However, the flatness in this region seems to better match a single linear behavior in the inertial domain of scales $\eta<l<\delta$ with slope $-0.1$ in agreement with the log-normal model of homogeneous and isotropic turbulence far from walls. Finally, we extended the study of structure functions to the streamwise velocity at positions $\tilde{y}$ in the viscous and buffer regions. In these regions, four different behaviors are found depending on the scale, with the inertial domain divided in two subdomains. However, now for scales $\eta<l<\delta$ the structure functions seems to scale following $\log(l/d)$ instead of $\log(l/\tilde{y})$. Also, no Kolmogorov scaling is found at the small scales of the inertial domain. Instead, we find equations (\ref{eq:Slog1visbuf}) and (\ref{eq:Slog2visbuf}) that illustrate the shear-dominated behavior of any scale in the inertial domain. The flatness in the buffer region reaches Gaussianity at scales smaller than $\delta$ with a decrease that could be correctly fitted by either equation (\ref{eq:FK41}) or (\ref{eq:Flog}). The flatness in the viscous region presents a steep linear decrease at small scales $\log(l/d)<0.5$ and converges to a plateau at $\log(F(l))=0.41$ for scales larger than $\log(l/d)=0.5$. This value of the flatness corresponds to a non-Gaussian distribution with heavy tails for the increments at these scales. In this region, where most of the energy is dissipated, the impact of the walls introduces a strong non-Gaussianity that is in agreement with strong dissipative structures: compared to a Gaussian, the distribution of the velocity increments presents heavier tails but also a higher bump at zero.

The ``attached eddy'' hypothesis of Townsend~\cite{Townsend1976}, that considers the presence of randomly distributed self-similar coherent structures whose size scales with $\tilde{y}$ in the logarithmic region, was used in previous works to explain eq.(\ref{eq:Slog})~\cite{Silva2015}. Moreover, numerical studies of turbulent channel flows revealed that above the buffer layer, the large scale structures with lengths of the order of $\mathcal{O}(3\tilde{y})$ cohabit and tend to align downstream, resulting in effectively much longer composite structures~\cite{Jimenez2013,Jimenez2013a}. This provides an alternative framework, not far from Townsend hypothesis, to explain the two inertial subdomains in the logarithmic region: at small scales $\eta<l<\tilde{y}$ the classical Kolmogorov scaling which comes from correlations inside the large scale structures, and at large scales $\tilde{y}<l<\delta$, the logarithmic scaling that results from the alignment of these large scale structures and from the correlations between them.

From our observations, the scaling of structure functions in both the viscous and buffer regions changes at scale $l \sim d$. For smaller scales, $\eta<l<d$, we find the same kind of scaling as for scales $\tilde{y}<l<\delta$ in the logarithmic layer, \textit{i.e.} these scales are dominated by the shear and the energy density scales as $1/l$ in this domain of scales. For larger scales, $d<l<\delta$, we find a different scaling with the fourth order structure function behaving linearly in $\log(l/d)$. Both scalings could be illustrative of different anisotropic turbulent structures oriented along the streamwise direction such as streaks~\cite{Cantwell1978} and quasi-streamwise vortices~\cite{Jimenez2013,Marusic2013a}. Thus, each scaling could correspond to different kind of structures, eq.(\ref{eq:Slog1visbuf}) for smaller ones and eq.(\ref{eq:Slog2visbuf}) for larger ones.

\acknowledgments

This work was supported by the French National Research Agency (ANR-21-CE46-0011-01), within the program ``Appel \`a projets g\'en\'erique 2021''.

\bibliographystyle{elsarticle-num} 
\bibliography{THEBIBLIO}

\end{document}